\documentclass{phbauth}
\usepackage{graphicx}

\begin{document}

\begin{frontmatter}

\title{Quasiparticle States near the Surface and the Domain Wall
in a $p_x$$\pm$i$p_y$-Wave Superconductor}

\author[address1]{Masashige Matsumoto\thanksref{thank1}},
\author[address2]{Manfred Sigrist}

\address[address1]{Department of Physics, Faculty of Science, Shizuoka University,
836 Oya, Shizuoka 422-8529, Japan}

\address[address2]{Yukawa Institute for Theoretical Physics, Kyoto University,
Kyoto 606-8502, Japan}


\thanks[thank1]{E-mail: spmmatu@ipc.shizuoka.ac.jp, Fax Number: 81-54-238-6352}

\begin{abstract}
The electronic states near a surface or a domain wall
in the $p_x$$\pm$i$p_y$-wave superconductor are studied.
This state has been recently suggested
as the superconducting state of Sr$_2$RuO$_4$.
The $p_x$$\pm$i$p_y$-wave paring state breaks the time reversal symmetry
and induces a magnetic field.
The obtained temperature dependence of the magnetic field is consistent
with the observed $\mu$SR data.
\end{abstract}

\end{frontmatter}

\newcommand{\bk}{\mbox{\boldmath$k$}}
\newcommand{\bkf}{\mbox{\boldmath$k_{\rm F}$}}
\newcommand{\bskf}{\mbox{\footnotesize \boldmath $k_{\rm F}$}}
\newcommand{\bd}{\mbox{\boldmath$d$}}
\newcommand{\ri}{{\rm i}}
\newcommand{\vf}{v_{\rm F}}
\newcommand{\vfy}{{v_{\rm F}}_y}
\newcommand{\kfy}{{k_{\rm F}}_y}

Studying unconventional superconductors has become one of the most
attractive problems in recent condensed matter research.
They include the recently discovered Sr$_2$RuO$_4$ \cite{Maeno}.
Triplet pairing state of $\bd(\bk)$=$(k_x$$\pm$$\ri k_y){\hat z}$ is suggested
as the d-vector \cite{Sigrist}.
Tunneling conductance for such paring state has been examined
finding that the conductance peak features related to the bound states
are very sensitive to the angle of the incidence of the electron \cite{Honerkamp,Yamashiro}.
Recently we have studied quasiparticle properties at the surface or domain wall
and reported that the local density of states at the surface
is constant and does not show any peak-like or gap-like structure
within the superconducting energy gap at low temperatures.
While at the domain wall it is v-shaped and contains a small gap-like feature \cite{Matsumoto}.

\begin{figure}[t]
\begin{center}\leavevmode
\includegraphics[width=0.56\linewidth]{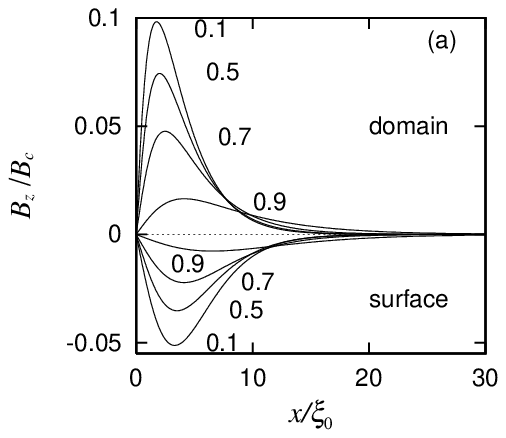}
\includegraphics[width=0.43\linewidth]{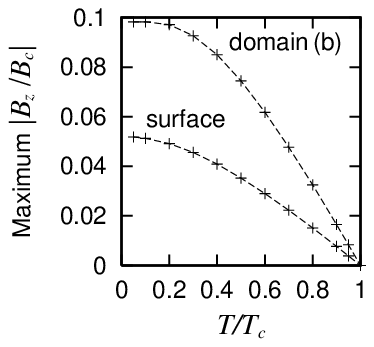}
\end{center}
\caption{Induced magnetic field.
$B_c$=$\Phi_0/2\sqrt{2}\pi\xi_0\lambda_{\rm L}$,
$\lambda_{\rm L}$ is the London penetration depth and $\Phi_0$=$h/2e$.
(a)~Spatial dependence at several temperatures.
$x$ is the distance from the surface or domain wall scaled by $\xi_0$=$\vf/\pi\Delta(0)$,
where $\Delta(0)$ is the magnitude of the bulk order parameter at $T$=0.
We chose a cutoff energy $\omega_c$=20$T_c$
and $\kappa$=$\lambda_{\rm L}/\xi_0$=2.5.
Temperatures scaled by $T_c$ are depicted.
$B_z$ is antisymmetric under $x$$\leftrightarrow$$-x$ for the domain wall.
(b)~Temperature dependence of the maximum $\mid B_z/B_c\mid$.}
\label{fig:1}
\end{figure}

The intrinsic magnetism in the superconducting phase by $\mu$SR experiment
indicates a pairing state with broken time reversal symmetry \cite{Luke}.
The magnetic field in the superconducting phase
can be induced by surface, domain wall and impurity \cite{Sigrist2}.
In this paper we examine the temperature dependence of the magnetic field
induced near the surface and domain wall
and compare them with the $\mu$SR experiment.
For this purpose we use the same formulation as in our previous paper \cite{Matsumoto},
which is based on the quasi-classical formulation developed by Schopohl $et$ $al$. \cite{Schopohl}.
The spatial variation of the order parameter and vector potential
can be determined self-consistently.
For simplicity we assume a two-dimensional $p_x$+i$p_y$ state.
In Fig. \ref{fig:1} we show the magnetic field near a surface and a domain wall
which is formed between $p_x-$i$p_y$ and $p_x$+i$p_y$ state.
Near the $T_c$ field maximum increases linearly with the decreasing the temperature
and it saturates at low temperatures.
This temperature dependence is qualitatively consistent with the $\mu$SR experiment.
In the surface case the energy level of the bound state is estimated as ${\Delta_y}(\bkf)$,
where ${\Delta_y}(\bkf)$ is the $p_y$-component of the order parameter with momentum $\bkf$.
Therefore, bound states in the region $\kfy$$<0$ are occupied
and yielding a spontaneous magnetic field,
as long as they satisfy the condition $T$$<$$\mid{\Delta_y}(\bkf)\mid$.

\begin{figure}[t]
\begin{center}\leavevmode
\includegraphics[width=0.5\linewidth]{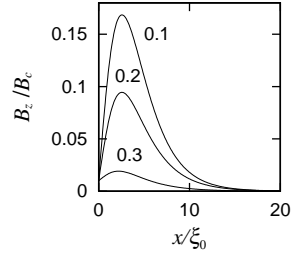}
\end{center}
\caption{Spatial dependence of the magnetic field
near the (1,0,0) surface for the $p_x$ state at several temperatures.
A small external field $B_{\rm ext}$=0.01$B_c$ is applied in the $z$ direction.
Temperatures scaled by $T_c$ are depicted.}
\label{fig:2}
\end{figure}

An interesting magnetic property appears in the case of a $p_x$ state.
It has been pointed out that midgap state gives rise to a paramagnetic response
\cite{Higashitani,Fogelstrom,Walter}.
Let us demonstrate it in the $p_x$ state,
which is suggested in the presence of the strong magnetic field
in the $x$ direction \cite{Agterberg}.
Figure \ref{fig:2} shows the spatial dependence of the paramagnetic field in the $z$ direction.
As it is studied in the $d$-wave case
energy level of the midgap state shifts to $e\vfy A_y$,
which splits the zero bias conductance peak \cite{Fogelstrom}.
Here $\vfy$ and $A_y$ are the $y$-component of Fermi velocity and vector potential, respectively.
Note that $A_y$ has the opposite sign of $B_{\rm ext}$.
Therefore, bound states in $\kfy$$>$0 region are occupied for a positive $B_{\rm ext}$
and it generates a magnetic field parallel to $B_{\rm ext}$.
Bound states satisfying $T$$<$$\mid e\vfy A_y\mid$ contribute to the effect
and the paramagnetic field rapidly decreases with the increase of temperature.
In the real case,
a small $p_y$ part can be induced by $B_{\rm ext}$.
The realized phase of the $p_y$ component is such that
generates a surface current which induces a filed parallel to $B_{\rm ext}$.
This results in also the paramagnetic response.
Without the strong field in the $x$ direction,
it is difficult to see this paramagnetic response,
since the occupied bound states are already asymmetric under $\kfy$$\rightarrow$$-\kfy$
and the state is difficult to modify with a small external field in the $z$ direction.

This work was supported by Grant-in-Aid for Encouragement of Young Scientists
from Japan Society for the Promotion of Science.


\end{document}